О.А. Юпиков, аспирант
Севастопольский национальный технический университет, Севастополь;
М.В. Ивашина, ст. научный сотрудник
The Netherlands Institute for Radio Astronomy (ASTRON), Dwingeloo, the Netherlands.

## ОПРЕДЕЛЕНИЕ ЧУВСТВИТЕЛЬНОСТИ МНОГОЛУЧЕВОЙ АНТЕННОЙ СИСТЕМЫ С ФОКАЛЬНОЙ РЕШЕТКОЙ

В данной статье описывается модель фокальной антенной решетки из элементов Вивальди, возможный подход к моделированию антенной системы зеркало–решетка–малошумящие усилители, а также приведены результаты моделирования и измерения чувствительности прототипа такой системы, построенной в институте *Astron* (Нидерланды).

Фокальные решетки (ФР) имеют ряд преимуществ перед традиционными зеркальными антеннами с рупором в качестве облучателей. Так как ФР может быть построена с близко расположенными элементами (< 0,5 $\lambda$), то открывается возможность одновременного формирования близко расположенных лучей. Это приводит к увеличению поля обзора антенны без потери разрешающей способности, к улучшению непрерывности поля обзора по сравнению с однолучевой системой с рупором или группой рупоров в качестве облучателя. Недостатком такой схемы является сильное взаимное влияние элементов решетки друг на друга. Вторым преимуществом фокальных решеток по сравнению с рупорами является то, что они могут быть построены из более широкополосных элементов. Это является важным моментом для некоторых приложений. Например, в радиоастрономии от широкополосности системы зависит скорость обзора участка неба.

Модель системы с фокальной решеткой может быть представлена в виде, показанном на рисунке 1. Система состоит из двух частей: 1) зеркало с элементами решетки, включающими металлическую структуру, устройство ее питания и малошумящие усилители (МШУ); 2) формирователь луча, представляющий собой управляемые аттенюаторы, фазовращатели и сумматор.

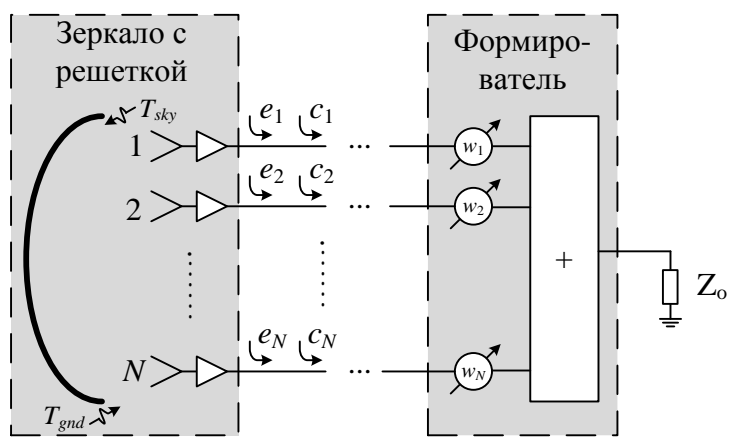

Рисунок 1 – Структура антенной системы

Несмотря на то, что каждая из этих подсистем имеет сложную структуру, и содержит внутренние источники сигнала и шума, каждая из них может быть описана матрицами рассеяния **S**, сигнальными волнами $e_n$, представленными в виде вектора **e**, и шумовой корреляционной матрицей **C**, элементами которой являются корреляционные коэффициенты между шумовыми волнами $c_n$. Поиск **S**, **e** и **C** для каждого блока делается предварительно путем совмещенного электромагнитного и микроволнового моделирования (см., например, [2, 3, 4, 5]), и не рассматривается в данной статье. Таким образом, при решении задачи оптимизации коэффициентов возбуждения может быть исключен анализ внутренних шумов и источников сигнала, генерируемых левой частью антенной системы (рисунок 1).

Для оптимизации весовых коэффициентов использован метод, предложенный в [1]. Максимальную чувствительность $(A_{eff}/T_{sys})_{max}$ можно выразить через отношение сигнал/шум $(SNR)_{max}$ по следующей формуле:

$$\left(A_{eff}/T_{sys}\right)_{max} = \left(2k_b/10^{-26}S_{src}\right)(SNR)_{max}, \qquad (1)$$

где $A_{eff}$ — эффективная площадь антенной системы; $T_{sys}$ — общая шумовая температура системы, включающая как внутренние шумы, так и принимаемые шумы неба и земли; $S_{src}$ — плотность потока энергии полезного источника в Янских; $k_b$ — постоянная Больцмана; $(SNR)_{max}$ — максимально достижимое отношение сигнал/шум данной системы при приеме сигнала данного источника, которое можно вычислить по следующему матричному выражению:

$$(SNR)_{max} = \mathbf{e}^H \mathbf{C}^{-1} \mathbf{e}, \qquad (2)$$

где **e** — сигнальный вектор, **C** — шумовая корреляционная матрица, верхний индекс $H$ обозначает эрмитов оператор. Зная **e** и **C**, вычисляются весовые коэффициенты для формирователя луча, оптимальные по критерию сигнал/шум, а значит, по критерию чувствительности:

$$\mathbf{w}_{max} = \mathbf{C}^{-1}\mathbf{e}. \qquad (3)$$

В рассматриваемом прототипе измерения вектора **e**, шумовых волн $c_n$, корреляционные коэффициенты между которыми составляют матрицу **C** (рисунок 1), проводились на одной из антенн Вестерборгского радиоинтерферометра (Нидерланды) следующим образом. Сначала телескоп направлялся на мощный астрономический источник с известной плотностью потока энергии $S_{src}$, и при помощи коррелятора измерялась корреляционная матрица сигнала **C** со всех 56 каналов решетки. Этот сигнал содержит как полезную составляющую источника, так и шумы системы (в основном МШУ) и окружения (шумы неба и земли). Далее телескоп направлялся на пустой участок неба, и снова измерялась корреляционная матрица сигналов $\mathbf{C}_0$ со всех каналов решетки, в этом случае сигналы содержат только шумы. После этого вычислялся сигнальный вектор как доминирующий собственный вектор разности $(\mathbf{C}-\mathbf{C}_0)$ [7]. Наконец, по формулам (1) — (3) вычислялись весовые коэффициенты для формирователя и чувствительность системы $(A_{eff}/T_{sys})_{max}$ для всех 17 направлений наблюдения. Наблюдения в данной системе проводятся по все направлениям одновременно, т.е. без перемещения антенны. Направленные лучи получаются после обработки полученных данных при помощи программного формирователя.

Моделирование проводилось для одних и тех же 17 наборов весовых коэффициентов, рассчитанных по выше приведенной методике для 17 направлений. Сопротивления элементов решетки и их диаграммы направленности моделировались с помощью расширенного метода моментов, известного как метод характеристических базисных функций [2], а потом объединялись с моделью устройства питания элементов [7]. Поверхностные токи и ДН элементов в масштабе, соответствующем весовым коэффициентам для осевого луча, показаны на рисунке 2.

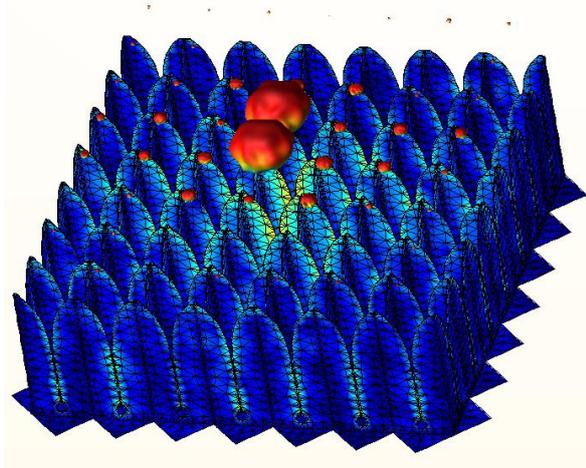

Рисунок 2 — Токи антенной решетки и ДН ее элементов с учетом весовых коэффициентов для центрального луча на частоте 1,42 ГГц

Для каждого из 17 направлений рассчитывалась общая ДН фокальной решетки $F_{FPA}$ как сумма ДН ее элементов $f_n$, умноженных на соответствующие весовые коэффициенты $w$:

$$F_{FPA}(\theta,\varphi) = \sum_{n=1}^{N} f_n(\theta,\varphi) w_n. \qquad (4)$$

После этого полученные первичные общие ДН передавались в программу GRASP9.3, в которой рассчитывались ДН после отражения от зеркала (вторичные ДН). Вторичные ДН использовались для нахождения усиления $G$ и, далее, эффективной площади антенной системы $A_{eff}$. Шумовая температура системы $T_{sys}$ анализировалась по методике, описанной в [4, 5]. После объединения всех полученных по этой методике шумовых составляющих и составляющих, учитывающих шумы неба и земли, получили общую шумовую температуру системы $T_{sys}$, и рассчитали чувствительность ($A_{eff}/T_{sys}$).

Некоторые рассчитанные коэффициенты эффективности антенной системы, ее шумовая температура и общая чувствительность в зависимости от направления луча, показаны соответственно на рисунках 3, 4 и 5. На этих рисунках использованы следующие обозначения: $\eta_{ap}$ — апертурный коэффициент использования поверхности (КИП) раскрыва зеркала, $\eta_{ill}$ — коэффициент эффективности амплитудно-фазового распределения в раскрыве зеркала, $\eta_{sp}$ — коэффициент перехвата энергии облучателя зеркалом, $T_{sp}$ — составляющая шумовой температуры системы за счет приема шумов земли, $T_{coup}$ — составляющая за счет потерь из-за сильного взаимного влияния элементов решетки друг на друга, $T_{min}$ — шумовой параметр МШУ (на данных графиках сюда так же включены шумы цепей, следующих за МШУ), $T_{sys}$ — общая шумовая температура системы.

Рассчитанный КИП $\eta_{ap}$, показанный на рисунке 3, изменяется в диапазоне (60…70) % при сканировании на ±1,2 градуса внутри поля обзора и резко уменьшается до 40 % при увеличении угла до 2 градусов. Такое поведение объясняется ограниченным размером фокальной решетки, т.к. основная часть энергии краевого луча попадает на краевые элементы. Высокая рассчитанная шумовая температура (100…130) К обусловлена, в основном, используемыми в прототипе МШУ. Однако, последняя разработка МШУ позволила уменьшить $T_{min}$ до 40 К [8]. Так же разрабатывается новая решетка с несколько большей активной площадью, что в совокупности с уменьшением $T_{min}$ даст увеличение чувствительности, как ожидается, примерно в 2 раза.

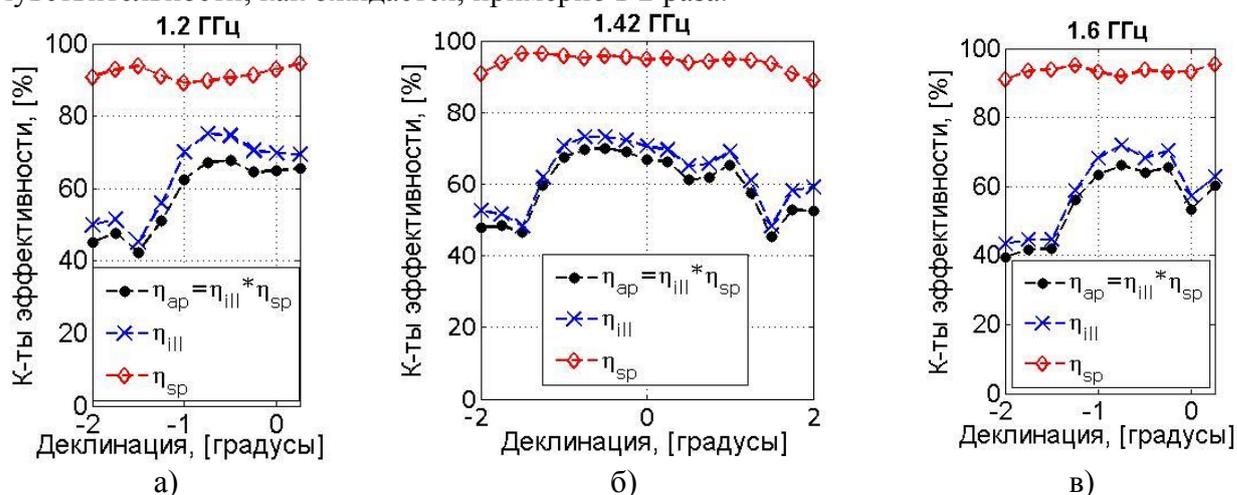

Рисунок 3 — Коэффициенты эффективности антенной системы, рассчитанные на частотах 1,2 ГГц (а), 1,42 ГГц (б) и 1,6 ГГц (в)

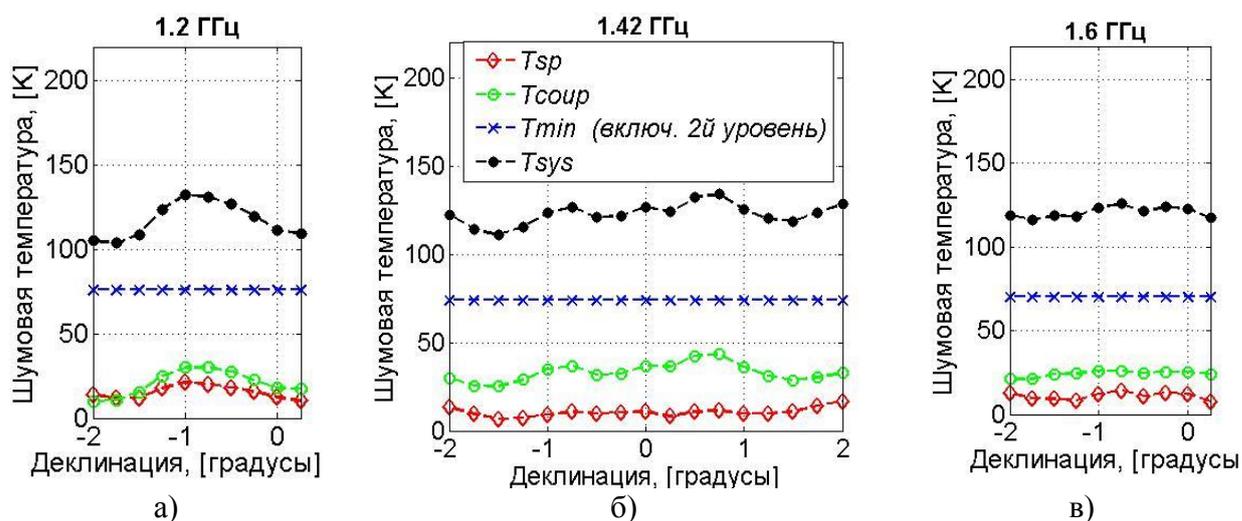

Рисунок 4 — Шумовая температура системы и ее основные составляющие, рассчитанные на частотах 1,2 ГГц (а), 1,42 ГГц (б) и 1,6 ГГц (в)

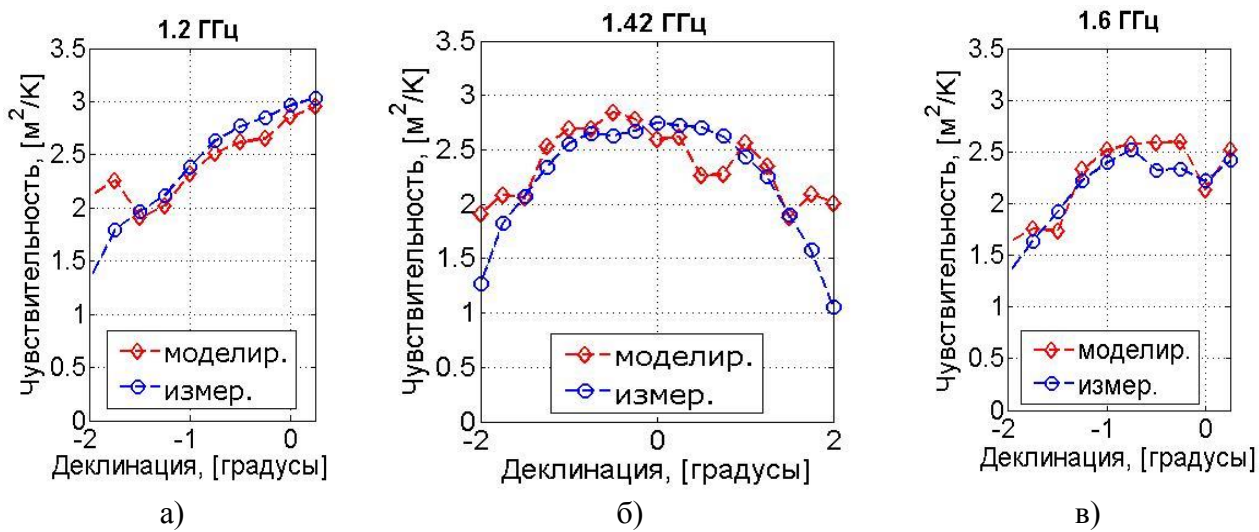

Рисунок 5 — Измеренная и промоделированная чувствительность антенной системы, полученные на частотах 1,2 ГГц (а), 1,42 ГГц (б) и 1,6 ГГц (в)

Таким образом, показано, что, во-первых, для увеличения равномерности распределения чувствительности внутри поля обзора необходимо увеличение площади фокальной решетки по сравнению с рассматриваемым прототипом, и, во-вторых, адекватность модели подтверждается результатами измерений.

В дальнейших исследованиях планируется рассмотреть новый прототип фокальной решетки, так как текущий имел несколько механических недостатков, которые приводили к нестабильности измерений как S-параметров решетки, так и корреляционных матриц, а также к их несимметричности для X и Y поляризованных элементов.

Также планируется ввести в модель эффект блокировки сигнала облучателем и держателями, а также провести анализ поляризационных свойств полученной системы.

Так как для радиоастрономии является важным, чтобы чувствительность в поле обзора была постоянна при ограниченном количестве одновременных лучей, то весовые коэффициенты необходимо будет оптимизировать с учетом этого требования. Это также является следующей задачей.

*Список литературы*

Acknowledgement
This work has been supported by the Netherlands Institute for Radio Astronomy ASTRON, and conducted during Iupikov's visit to ASTRON during 2009 under the supervision of Drs. Ivashina, Maaskant and Cappellen.